\documentclass[prl,twocolumn,10pt,a4paper,superscriptaddress,showpacs]{revtex4}
\usepackage{amsmath,epsfig}

\def\opone{\leavevmode\hbox{\small1\kern-3.8pt\normalsize1}}
\input{tcilatex}

\begin{document}

\title{An experimental observation of geometric phases for mixed states
using NMR interferometry}
\author{Jiangfeng \surname{Du}}
\email{djf@ustc.edu.cn}
\affiliation{Structural Research Laboratory and Department of Modern Physics, University
of Science and Technology of China, Hefei, 230027, P.R.China.}
\affiliation{Department of Physics, National University of Singapore, 2 Science Drive 3,
Singapore 117542.}
\affiliation{Centre for Quantum Computation, DAMTP, University of Cambridge, Wilberforce
Road, Cambridge CB3 0WA , U.K.}
\author{Ping Zou}
\affiliation{Structural Research Laboratory and Department of Modern Physics, University
of Science and Technology of China, Hefei, 230027, P.R.China.}
\author{Mingjun \surname{Shi}}
\affiliation{Structural Research Laboratory and Department of Modern Physics, University
of Science and Technology of China, Hefei, 230027, P.R.China.}
\author{Leong Chuan \surname{Kwek}}
\affiliation{Department of Natural Sciences, National Institute of Education, Nanyang
Technological University, 1 Nanyang Walk, Singapore 637616.}
\author{Jian-Wei \surname{Pan}}
\affiliation{Structural Research Laboratory and Department of Modern Physics, University
of Science and Technology of China, Hefei, 230027, P.R.China.}
\author{Choo Hiap \surname{Oh}}
\affiliation{Department of Physics, National University of Singapore, 2 Science Drive 3,
Singapore 117542.}
\author{Artur \surname{Ekert}}
\affiliation{Centre for Quantum Computation, DAMTP, University of Cambridge, Wilberforce
Road, Cambridge CB3 0WA , U.K.}
\affiliation{Department of Physics, National University of Singapore, 2 Science Drive 3,
Singapore 117542.}
\author{Daniel K. L. \surname{Oi}}
\affiliation{Centre for Quantum Computation, DAMTP, University of Cambridge, Wilberforce
Road, Cambridge CB3 0WA , U.K.}
\author{Marie \surname{Ericsson}}
\affiliation{Dept. of Physics, University of Illinois at Urbana-Champaign, Urbana IL
61801-3080, United States of America.}

\begin{abstract}
Examples of geometric phases abound in many areas of physics. They offer
both fundamental insights into many physical phenomena and lead to
interesting practical implementations. One of them, as indicated recently,
might be an inherently fault-tolerant quantum computation. This, however,
requires to deal with geometric phases in the presence of noise and
interactions between different physical subsystems. Despite the wealth of
literature on the subject of geometric phases very little is known about
this very important case. Here we report the first experimental study of
geometric phases for mixed quantum states. We show how different they are
from the well understood, noiseless, pure-state case.
\end{abstract}

\maketitle

A quantum system can retain a memory of its motion when it undergoes a
cyclic evolution, e.g its quantum state may acquire a geometric phase factor
in addition to the dynamical one~\cite{Panchar56,Ber84}. For pure quantum
states this effect is well understood and it has been demonstrated in a wide
variety of physical systems~\cite{GPP89}. Its potential application to
perform the fault-tolerant quantum computation has been the subject of more
recent investigations~\cite{JVEC00,Zan99,DCZ01}. In contrast, relatively
little is known about geometric phases, and more generally, about quantum
holonomies of mixed or entangled quantum states. Here we report an NMR
experiment which constitutes the first experimental study of quantum
holonomies for mixed quantum states. We observed and measured the geometric
phase of a mixed state of a spin half nuclei. Our experimental data are in a
good agreement with the recent theoretical predictions by Sj\"{o}qvist et al~%
\cite{sjoqvist}.

The geometric phase of pure states is an intriguing property of quantum
systems undergoing parallel cyclic evolutions. The parallel transport of a
particular vector $|\Psi\rangle$ implies no change in phase when $%
|\Psi(t)\rangle$ evolves into $|\Psi(t+dt)\rangle$, for some infinitesimal
change of the parameter $t$. Although locally there is no phase change, the
system may acquire a non-trivial phase after completing a closed loop
parameterized by $t$. The origin of this phase can be traced to an
underlying curvature of the parameter space, depending only on the geometry
of the path and is resilient to certain dynamical perturbations of the
evolution, e.g. it is independent of the speed of the evolution. Therefore,
it is a potential method for performing intrinsically fault-tolerant quantum
logic gates, a very desirable feature for practical implementations of
quantum computation. However, quantum systems that interact with other
systems, be it components in a quantum computer or otherwise, become
entangled and cannot be described by a state vector $|\Psi\rangle$. In this
context the notion of parallel transport and geometric phases must be
extended to mixed quantum states.

Mathematically, Uhlmann was the first to address the issue of a mixed state
holonomy~ \cite{uhlmann}. In his approach a system in a mixed state is
embedded, as a subsystem, in a larger system that is in a pure state. Given
a mixed state of the subsystem there are infinitely many corresponding pure
states, known as purifications, of the larger system. Thus a cyclic
evolution of the density operator pertaining to the subsystem induces
infinitely many possible evolutions of the larger, purified system. Uhlmann
singles out the evolution in which the purified state is transported in a
maximally parallel manner. In order to satisfy this condition, one has to
induce a suitable evolution on all auxiliary subsystems with which the
original subsystem is entangled.

More recently Sj\"{o}qvist et al~\cite{sjoqvist} took a different approach
in which there is no need for a direct reference to auxiliary subsystems~%
\cite{EPSBO2002}. In their case each eigenvector of the initial density
matrix is parallel transported independently and may acquire a geometric
phase factor $\gamma_n$. The mixed state phase factor is then obtained as an
average of the individual phase factors, weighted by their eigenvalues $p_n$%
, 
\begin{equation}
v e^{i\gamma} = \sum_n p_n e^{i\gamma_n}.  \label{gemphase}
\end{equation}
This geometric phase factor can also be understood using purifications~\cite%
{sjoqvist}, though operations on auxiliary subsystems are unconstrained and
as such they include Uhlmann's approach as a special case.

Definitions, by definition, are never wrong or right, just more or less
useful, thus we are not in a position to refute either of the two
approaches. Here we investigate the holonomies of mixed quantum states and
show that experimental data are consistent with the approach by Sj\"{o}qvist
et al~\cite{sjoqvist}.

\begin{figure}[tbp]
\begin{center}
\epsfig{figure=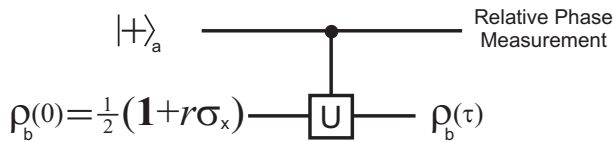,width=0.45\textwidth}
\end{center}
\caption{A quantum network describing the experiment. The top horizontal
line represents an auxiliary spin half particle, or an auxiliary qubit,
labelled as qubit ``a". The bottom line represents a qubit labelled as ``b",
in state $\protect\rho_b$ which undergoes a cyclic evolution induced by a
unitary operation $U$. We choose our reference basis, for qubits``a" and
``b", to be states $|\uparrow\rangle$ and $|\downarrow\rangle$. They
describe the spin state aligned with or against a static magnetic field $B_0$
applied in the $z$-direction. In this basis $|\pm\rangle=\frac{1}{\protect%
\sqrt 2}( |\uparrow\rangle\pm|\downarrow\rangle)$ thus the initial state of
the auxiliary qubit is $|+\rangle_a$. Projectors $|\pm\rangle\langle \pm|$
can also be written as $\mbox{$\textstyle \frac{1}{2}$}(1\pm\protect\sigma %
_{x})$.}
\label{fig:network}
\end{figure}

In our NMR experiment we focused on a mixed state of a spin half nuclei. Its
density operator can be written in terms of the Bloch vector $\vec r$ and
the Pauli matrices $\vec\sigma=\{\sigma_x, \sigma_y, \sigma_z\}$, as $\rho=%
\mbox{$\textstyle \frac{1}{2}$}(\leavevmode\hbox{\small1\kern-3.8pt%
\normalsize1}+\vec r\cdot \vec\sigma)$. It represents a mixture of its two
eigenvectors with eigenvalues $\mbox{$\textstyle \frac{1}{2}$} (1\pm r)$.
The length of the Bloch vector $r$ gives the measure of the purity of the
state -- from maximally mixed $r=0$ to pure $r=1$. We use the gradient
pulses to produce mixed states of different purities $r$. We then evolve the
Bloch vector $\vec r$ so that it traces out a curve $C$ that subtends the
solid angle $\Omega $. For spin half particles, Eq.~(\ref{gemphase}) gives 
\begin{equation}
v e^{i\gamma} = \cos\Omega/2 + i r\sin\Omega/2.  \label{average}
\end{equation}
This phase factor can be estimated from the visibility in an interference
experiment~\cite{sjoqvist}. In our experiment we measure the geometric phase 
\begin{equation}
\gamma=-\arctan \left( r\tan \frac{\Omega }{2}\right)
\end{equation}
using an auxiliary spin half particle for the phase reference. A succinct
description of the experiment is given in Fig.~(\ref{fig:network}). Adopting
the nomenclature from quantum information science we will sometimes refer to
spin half particles as qubits.

\begin{figure}[tbp]
\begin{center}
\epsfig{figure=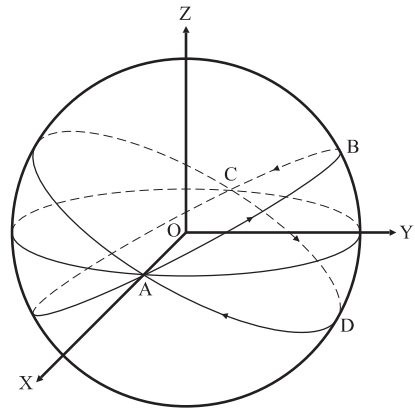,width=0.30\textwidth}
\end{center}
\caption{The cyclic path ABCDA subtends the solid angle $\Omega$ on the
Bloch sphere. The solid angle can be changed by varying $\protect\theta$ -
the angle of inclination between the $x,y$-plane and the ABC plane (or the
ADC plane). In the experiment the Bloch vectors of different lengths $r$
follow the path identical to the ABCDA but right below it, at the distance $%
r $ from the centre of the sphere. The two Bloch vectors corresponding to
the two eigenvectors of the density operator are the unit vectors $\pm \vec
r/r$. In the cyclic evolution one of them follows the path ABCDA and
subtends the solid angle $\Omega$ and the other follows the path
symmetrically on the opposite side of the sphere and subtends the solid
angle $-\Omega$.}
\label{fig:path}
\end{figure}

The central element in Fig.~(\ref{fig:network}) is the controlled-$U$
operation. In our case: the state $\rho_b$ traces out a closed path $C:t\in %
\left[ 0,\tau \right] \rightarrow \rho\left( t\right)=U\left( t\right)
\rho\left( 0\right) U^{\dag }\left( t\right) $ on the Bloch sphere with a
solid angle $\Omega$, but only when the auxiliary qubit is in state $%
|\uparrow\rangle_a$; when the auxiliary qubit is in state $%
|\downarrow\rangle_a$ the state $\rho_b$ is not affected. Such a controlled
evolution can be realized in NMR with the scalar spin-spin coupling of the
two spins. It effectively introduces a relative phase shift between the
states $|\uparrow\rangle_a$ and $|\downarrow\rangle_a$ of the auxiliary
qubit. In the experiment, the unitary operation which induces the cyclic
motion completes the loop along the two geodesics, ABC and CDA, as
illustrated in Fig.(\ref{fig:path}). It satisfies the parallel transport
condition defined in~\cite{sjoqvist} and thus the dynamic phase vanishes.

In our experiment the geometric phase for mixed states was observed using an
NMR spectrometer. We used a 0.5 ml, 200 mmol sample of Carbon-13 labelled
chloroform (Cambridge Isotopes) in d$_{6}$ acetone. The single $^{13}C$\
nucleus was used as the auxiliary qubit while the $^{1}H$\ nucleus was used
as the qubit on which the cyclic evolution was executed. The reduced
Hamiltonian for this two-spin system is, to an excellent approximation,
given by 
\begin{equation}
H=\omega _{a}I_{z}^{a}+\omega _{b}I_{z}^{b}+2\pi J\;I_{z}^{a}I_{z}^{b}.
\end{equation}
The first two terms in the Hamiltonian describe the free precession of spin
``a" ($^{13}C$) and spin ``b" ($^{1}H$) around the magnetic field $B_{0}$
with frequencies $\omega _{a}/2\pi \approx 100$MHz and $\omega _{b}/2\pi
\approx 400$MHz. The $I_{z}^{a}$ and $I_{z}^{b}$ are the $z$-components of
the angular momentum operator for ``a" and ``b" respectively ($I_{z}\equiv%
\mbox{$\textstyle \frac{1}{2}$}\sigma_z$). The third term of the Hamiltonian
describes a scalar spin-spin coupling of the two spins with $J=214.5$Hz. In
our experiment, we varied the solid angle $\Omega$, and for each $\Omega $
we measured the geometric phase $\gamma$ for twelve values of $r=\cos \frac{%
n\pi }{12},\; n=0,1,\ldots,11$.

Let us now describe step by step different stages of the experiment in more
detail.

(E1) \textbf{Preparation of the initial state:} Initially the two qubits are
in thermal equilibrium with the environment and their state is described by
the density operator $\rho _{th}\propto \sigma_{z}^{a}+4\sigma_{z}^{b}$. We
use the spatial averaging technique~\cite{cory} to create the effective pure
state $|\uparrow\rangle_a\otimes|\uparrow\rangle_b$ or, in the density
operator form, $\mbox{$\textstyle \frac{1}{2}$}(1+\sigma _{z}^a) \otimes %
\mbox{$\textstyle \frac{1}{2}$}(1+\sigma _{z}^b)$. The sequence of
operations leading to this state, reading from the left to the right, is as
follows, 
\begin{equation}
R_{x}^{b}\left( \pi /3\right) -G_{z}-R_{x}^{b}\left( \pi /4\right) -\frac{1}{%
2J} -R_{-y}^{b}\left( \pi /4\right) -G_{z},
\end{equation}
where $R_{x}^{b}(\alpha) =e^{-i\alpha\sigma _{x}/2}$ denotes a selective
pulse that rotates the spin $b$ around the $x$-axis by angle $\alpha $ (and $%
R_{-x}^{b}(\alpha)\equiv R_{x}^{b}(-\alpha)$), $G_{z}$ is the pulsed field
gradient along the $z$-axis (it annihilates the transverse magnetizations),
and $\frac{1}{2J}$ represents just a time interval of $1/\left(2J\right)$.
Note that the above pulse sequence is different from the one described in~%
\cite{cory} because we used a heteronuclei rather than a homonuclei sample.
The subsequent pulse sequence 
\begin{equation}
R_{x}^{b}\left( n\pi /12\right) -G_{z}-R_{-y}^{a}\left( \pi /2\right)
-R_{-y}^{b}\left( \pi /2\right)
\end{equation}
generates the desired initial state 
\begin{equation}
\rho _{ab}\left( 0\right) \equiv \rho _{a}\left( 0\right) \otimes \rho
_{b}\left( 0\right) =\mbox{$\textstyle \frac{1}{2}$}(\leavevmode%
\hbox{\small1\kern-3.8pt\normalsize1}+\sigma _{x}^a)\otimes %
\mbox{$\textstyle \frac{1}{2}$}(\leavevmode\hbox{\small1\kern-3.8pt%
\normalsize1}+r\sigma _{x}^b)
\end{equation}
with purity $r=\cos (n \pi/12);\; n=0,1,\cdots,11$, which is set by the
rotation angle $n\pi/12$ of the selective pulse $R_{x}^{b}(n\pi/12)$.

(E2) \textbf{The controlled-$U$ operation:} This operation is implemented
setting the oscillation frequency $\omega_{b}^{\prime}=\omega _{b}-\pi J$,
so that the Hamiltonian of qubit $b$ in the rotating frame with angular
frequency $\omega_{b}^{\prime}$ can be written as $H_{b}(0)
=(\omega_{b}-\omega_{b}^{\prime}\pm\pi J)\; I_{z}^{b}$. The $\pm$ sign is
determined by the state of qubit $a$. If qubit $a$ is in state $%
|\uparrow\rangle_a$ then $H_{b}(0)=0$; if qubit $a$ is in state $%
|\downarrow\rangle_a$ then $H_{b}(0)=2\pi J\;I_{z}^{b}$. Subsequently we use
the following pulse sequence to implement the cyclic evolution, 
\begin{equation}
R_{-x}^{b}\left( \theta \right) -\frac{1}{2J}-R_{-x}^{b}\left( \pi -2\theta
\right) -\frac{1}{2J},
\end{equation}
where $\theta =\Omega /4$ is the inclination angle (see Fig.(\ref{fig:path}%
)). The effect of this evolution is illustrated in Fig.(\ref{fig:path}). The
two eigenstates of $\rho_b(0)$, namely $|\pm\rangle_b$, trace out a path
which encompasses the solid angle $\Omega$ and acquire geometric phases $%
|\pm\rangle_b\mapsto e^{\mp i\frac{\Omega }{2}}|\pm\rangle_b$. Since the
path follows geodesics, the dynamical phase disappears. Thus the auxiliary
qubit acquires the phase factor $e^{\mp i\frac{\Omega }{2}}$, i.e. $\frac{1}{%
\sqrt 2}( |\uparrow\rangle+|\downarrow\rangle)\mapsto \frac{1}{\sqrt 2}(
|\uparrow\rangle+e^{\mp i\frac{\Omega }{2}}|\downarrow\rangle)$, with the
probability $\mbox{$\textstyle \frac{1}{2}$}(1\pm r)$. Their averaging gives
the phase factor as in Eq.(\ref{average}).

(E3) \textbf{Measurement:} In the experiment we use a phase sensitive
detector to measure the phase $\gamma$ relative to the reference phase of
the initial state $|+\rangle_a$.

\begin{figure}[tbp]
\begin{center}
\epsfig{figure=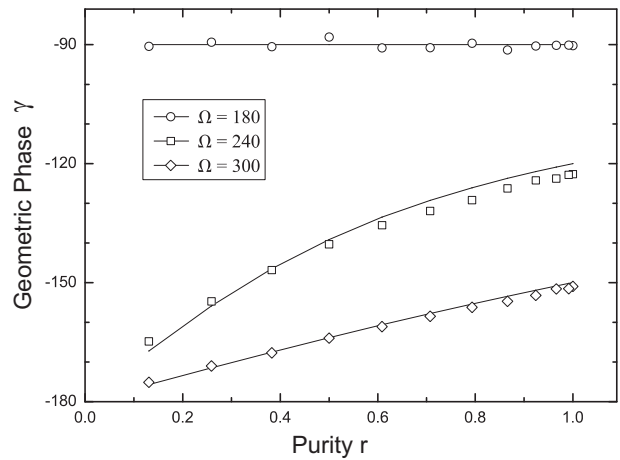,width=0.45\textwidth}
\end{center}
\caption{Summary of experimentally determined geometric phase $\protect%
\gamma $ as a function of purity of the mixed state for three different
solid angles $\Omega$. The solid lines correspond to the theoretical result: 
$\protect\gamma =-\arctan \left( r\tan \frac{\Omega }{2}\right) $.}
\label{fig:result}
\end{figure}

Fig. \ref{fig:result} shows a plot of $\gamma$ versus the purity of mixed
state for the three different solid angles $\Omega $. The experimental data
and the theoretical prediction are in a very good agreement. The small
errors are due to inhomogeneity of magnetic field and imperfect pulses.

All experiments were conducted at room temperature and pressure on Bruker
AV-400 spectrometer. In our experiment, all the pulses are square and are of
several microseconds duration. The spin-spin relaxation times are $0.3$s for
carbon and $0.4$s for proton, respectively. In each experiment, the time
used for the cyclic parallel transport evolution is about $4.7$ms, which is
well within the decoherence time.

To summarize, we have experimentally observed geometric phases for mixed
states which are in accordance with the theoretical predictions. In the
future, we should be able to extend the study of geometric phases to the
non-unitary regime which is especially pertinent to their application to
fault tolerant quantum computation~\cite{ESBOP2002}.

\begin{acknowledgments}
We thank Zeng-Bing Chen, Y.-D. Zhang, M.V. Berry, V. Vedral and Jihui Wu for
helpful discussions. This project was supported by the National Nature
Science Foundation of China (Grants. No. 10075041 and No. 10075044) and
Funded by the National Fundamental Research Program (2001CB309300) and the
ASTAR Grant No. 012-104-0040. DKLO acknowledges support from EU grant TOPQIP
(IST-2001-39215) and the CMI project in Quantum Information Science. ME
acknowledges support of the Foundation BLANCEFLOR Boncompagni-Ludovisi, ne%
\'{e} Bildt.
\end{acknowledgments}

\end{document}